\documentclass[11pt]{revtex4}

\usepackage{epsfig}

\begin{document}

\title{Superlattice with hot electron injection: an approach to a Bloch oscillator}

\author{D.A. Ryndyk$^{1,3}$, N.V. Demarina$^{2,4}$, J. Keller$^1$, and E. Schomburg$^2$}

\affiliation {$^1$Institute for Theoretical Physics, University of Regensburg,
Germany \\ $^2$Institute for Experimental and Applied Physics, University of Regensburg,
Germany \\ $^{3}$Institute for Physics of Microstructures RAS, Nizhny Novgorod,
Russia \\ $^{4}$Department of Radiophysics, University of Nizhny Novgorod,
Russia}

\begin{abstract}
A semiconductor superlattice with {\em hot electron injection into the
miniband} is considered. The injection changes the stationary distribution
function and results in a qualitative change of the frequency behaviour of the
differential conductivity. In the regime with Bloch oscillating electrons and
{\em injection into the upper part of the miniband} the region of negative
differential conductivity is shifted from low frequencies to higher
frequencies. We find that the {\em dc} differential conductivity can be made
positive and thus the domain instability can be suppressed. At the same time
the {\em high-frequency} differential conductivity is negative {\em above} the
Bloch frequency. This opens a new way to make a Bloch oscillator operating at
THz frequencies.
\end{abstract}

\date{\today}

\maketitle

Miniband electron transport in semiconductor superlattices is an important
subject of experimental and theoretical investigations beginning from the
pioneering paper of Esaki and Tsu \cite{Esaki70}. If a static electric field
$E_s$ is applied along the superlattice axes, electrons begin to move in
accordance with the semiclassical Newton's law (neglecting scattering)
\begin{equation}\label{Newton's law}
  \frac{dp}{dt}=eE_s,
\end{equation}
where $p$ is the momentum along the superlattice axes. If the energy gap
$\Delta_G$ between the first and the second miniband is large enough,
$\Delta_G\gg edE_s$ ($d$ is the superlattice period) and the scattering rate
$\nu$ is small, $h\nu<edE_s$, then electrons oscillate inside the first
miniband with the so-called Bloch frequency $\Omega=edE_s/\hbar$. The
quasiparticle energy $\epsilon_p=(\Delta/2)(1-\cos(dp/\hbar))$ and the
quasiparticle group velocity along the superlattice axes
$v_p=\partial\epsilon_p/\partial p =(\Delta d/2\hbar) \sin(dp/\hbar)$ are
periodic functions of time ($\Delta$ is the miniband width). We assume that the
condition of semiclassical approximation $\Delta\gg edE_s$ is fulfilled.

\begin{figure}
  \centering
  \begin{minipage}{0.4\textwidth}
  \epsfxsize=\hsize
  \epsfbox{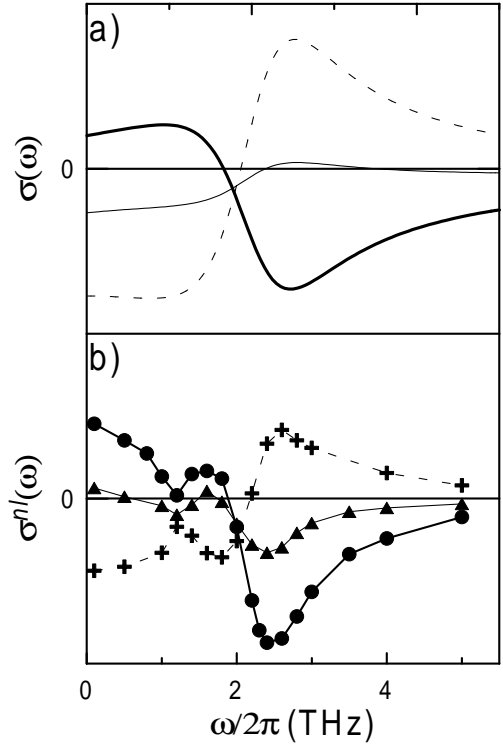}
  \end{minipage}
  \caption{Differential conductivities (a) analytical results:
  $\sigma_E(\omega)$ (broken line);
  $\sigma(\omega)$ at $\eta=0.036$ (thin line) and $\eta=0.072$ (thick line);
  (b) 3D Monte-Carlo simulation: $\sigma^{nl}_E(\omega)$ (crosses);
  $\sigma^{nl}(\omega)$ at $\eta=0.036$ (triangles) and $\eta=0.072$ (circles);
  $\nu/\Omega=0.36$, $\Delta/2T=0.3$, $\varphi'=0.9\pi$.}
  \label{sigma-fig}
\end{figure}

In semiconductor superlattices the Bloch frequency can be varied by the
electric field up to 1-10 THz, so it is a great challenge to make a tunable
Bloch oscillator in the THz frequency range. However, in most cases the phases
of Bloch oscillating electrons are random and high-frequency emission is
noise-like. The amplification of an external high-frequency signal is possible
due to a phase bunching of electrons. In 1972 S.A. Ktitorov, G.S. Simin, and
V.Ya. Sindalovskii \cite{Ktitorov72} showed that the linear differential
conductivity $\sigma_E(\omega)$ of a biased superlattice is negative up to the
Bloch frequency (see Fig.\,\ref{sigma-fig}(a), broken line). Later A.A.
Ignatov, K.F. Renk, and E.P. Dodin \cite{Ignatov93} calculated the finite-field
ac response of a biased superlattice and showed that the amplification
efficiency, i.e. the ratio of power absorbed by a THz field and the power
delivered from a bias source, is larger at larger ac fields. That is what one
needs to make an oscillator work at $\omega\sim\Omega$. However,
$\sigma_E(\omega)$ is negative also at $\omega\rightarrow 0$, and this dc
negative differential conductivity (NDC) leads to a low-frequency instability
of space-charge waves and to the formation of domains
\cite{Buttiker77-79,Schomburg98}. This effect can be used itself for a
generation of microwaves by moving domains with frequencies above 100 GHz
\cite{Schomburg99}, but the efficiency of a THz radiation at Bloch frequencies
cannot be large in the inhomogeneous regime. {\em For a Bloch oscillator to
operate the domain instability should be suppressed}. It was shown by Yu.A.
Romanov, V.P. Bovin, and L.K. Orlov \cite{Romanov78} and by H. Kroemer
\cite{Kroemer00} that in the {\em nonlinear} regime with high enough
high-frequency current the dc differential conductivity is {\em positive},
while the large-signal high-frequency differential conductivity still remains
to be negative, consequently, the steady-state operation of a Bloch oscillator
in large-signal regime can be achieved. However, the question of "device
turn-on" is not answered. Note also, that the nonlinear regime can be unstable
\cite{Romanov00}.

We present here a new approach to this long-standing problem. The key idea is
to inject electrons into the upper part of the first miniband (note the
discussion of other new effects in a superlattice with hot electrons in
\cite{Cannon00}). We calculate the nonequilibrium stationary distribution
function, and both {\em linear} and {\em nonlinear} (large ac current)
differential conductivities with the help of a one-dimensional kinetic equation
and 3D Monte-Carlo simulations. We show that the low-frequency domain
instability is suppressed and the region of negative differential conductivity
is shifted from low frequencies to higher frequencies. Thus the self excitation
of high-frequency oscillations becomes possible. At a finite level of
high-frequency current this effect is supported by dynamic domain suppression
found by Romanov and Kroemer.

We restrict our analytical consideration to the one-dimensional problem, where
the dynamics of the system can be described by a semiclassical Boltzmann
equation for the distribution function $f(p,x,t)$.  The distribution function
determines the particle density
\begin{equation}\label{density}
  n(x,t)=\int_{-\pi\hbar/d}^{\pi\hbar/d} f(x,p,t)dp,
\end{equation}
and the current density
\begin{equation}\label{current}
  j(x,t)=e\int_{-\pi\hbar/d}^{\pi\hbar/d} v_pf(x,p,t)dp.
\end{equation}

In a one-miniband approximation with a Bhatnagar-Gross-Krook (BGK) scattering
integral, which conserves the number of particles and makes it possible to
describe space-charge effects \cite{Ignatov87}, we have
\begin{equation}\label{Boltzmann equation}
  \frac{\partial f}{\partial t}+v_p\frac{\partial f}{\partial x} +eE\frac{\partial f}{\partial p}
  =-\nu\left(f-\frac{n}{n_0}f_0\right)+S(p),
\end{equation}
where
\begin{equation}\label{f_0}
  f_0=\frac{n_0d}{2\pi\hbar
  I_0(\Delta/2T)}\exp\left(\frac{\Delta}{2T}\cos\frac{dp}{\hbar}\right)
\end{equation}
is the equilibrium distribution function of a non-degenerate electron gas  and
$n_0$ is the equilibrium particle density. $S(p)$ is the source of hot
particles. For simplicity (and to conserve the total charge) we choose a
"narrow" source of the simplest form
\begin{equation}\label{in-delta}
S(p)=Q\delta(p-p')-\frac{Q}{n_0}f_s(p),
\end{equation}
where $f_s(p)$ is the stationary (static and homogeneous) solution of
(\ref{Boltzmann equation}). $Q$ is the injection rate of hot electrons and $p'$
is their momentum.

\begin{figure}
  \centering
  \begin{minipage}{0.4\textwidth}
  \epsfxsize=\hsize
  \epsfbox{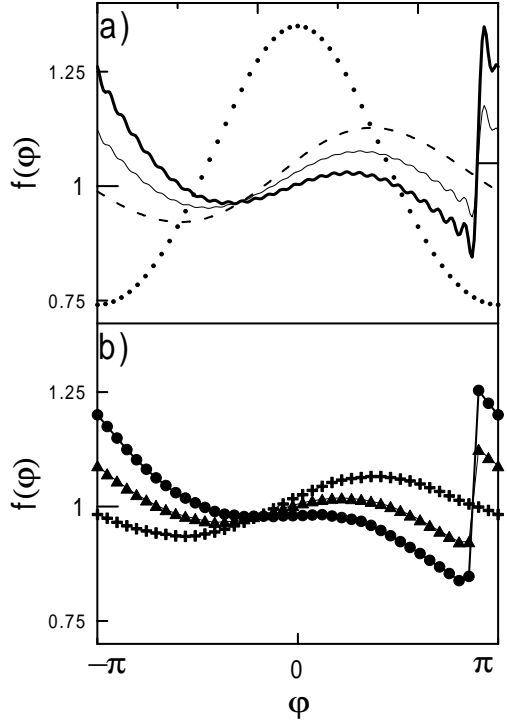}
  \end{minipage}
  \caption{Stationary distribution functions in the miniband: (a) analytical results for
  $f_0(\varphi)$ (dots); $f_E(\varphi)$ (broken line);
  $f_s(\varphi)$ at $\eta=0.036$ (thin line) and $\eta=0.072$ (thick line);
  (b) 3D Monte-Carlo simulation without injection (crosses);
  at $\eta=0.036$ (triangles) and $\eta=0.072$ (circles);
  $\nu/\Omega=0.36$, $\Delta/2T=0.3$, $\varphi'=0.9\pi$.}
  \label{df-fig}
\end{figure}

In the homogeneous case (constant electric field $E_s$) and $S(p)=0$ the
solution of Eq.\,(\ref{Boltzmann equation}) is
\begin{equation}\label{f_E}
  f_E(\varphi)=\frac{n_0d}{2\pi\hbar I_0\left(\frac{\Delta}{2T}\right)}\sum_l
  I_l\left(\frac{\Delta}{2T}\right)\frac{\nu}{\nu+il\Omega}e^{il\varphi},
\end{equation}
where $\varphi=dp/\hbar$ is the dimensionless momentum, $I_l$ is the modified
Bessel function of $l$-th order. Note, that due to the one-miniband
approximation $f(x,\varphi,t)$ is a periodic function of $\varphi$ with a
period of $2\pi$.

In presence of the hot electron source (\ref{in-delta}) the stationary
homogeneous distribution function $f_s(\varphi)$ can be represented as
$f_s=f_E+f'$ and $f'(\varphi)$ can be obtained from the equation
\begin{equation}
  \Omega\frac{\partial f'}{\partial\varphi}=-\left(\nu+\frac{Q}{n_0}\right)f'+
  \frac{dQ}{\hbar}\delta(\varphi-\varphi')-
  \frac{Q}{n_0}f_E(\varphi).
\end{equation}
Thus we find
\begin{equation}\label{f_s}
  f'(\varphi)=\eta\frac{n_0d}{2\pi\hbar}\sum_l\frac{\Omega e^{il\varphi}}{\nu+\eta\Omega+il\Omega}
  \left[e^{-il\varphi'}-\frac{I_l\left(\frac{\Delta}{2T}\right)}{I_0\left(\frac{\Delta}{2T}\right)}
  \frac{\nu}{\nu+il\Omega}\right],
\end{equation}
where we have introduced the parameter of nonequilibrium
\begin{equation}
\displaystyle \eta=\frac{Q}{\Omega n_0}.
\end{equation}

Distribution functions $f_s(\varphi)$ at different injection rates are shown in
Fig.\,\ref{df-fig}(a) for a superlattice \cite{Schomburg98} with $\Delta=16$
meV, $d=5.06$ nm, $\nu=0.5\cdot 10^{13} $ sec$^{-1}$, $E_s=18$ kV/cm,
$\Omega=1.38\cdot 10^{13}$ sec$^{-1}$, at room temperature. We also calculated
the distribution function by using a one-particle Monte-Carlo method
\cite{Jacoboni83}, where the motion of an electron described by the
Eq.\,(\ref{Newton's law}) is interrupted by three-dimensional scattering events
(on acoustical and optical phonons), randomly distributed in time with the
average frequency $\nu$. The averaged frequency for each scattering process was
calculated from quantum mechanical perturbation theory taking into account the
superlattice energy dependence and the material parameters of bulk GaAs. The
injection is modeled in the same manner by the random change of the momentum to
$p'$ with the averaged frequency $Q/n_0$. The results of our Monte-Carlo
simulations are presented in Fig.\,\ref{df-fig}(b). The comparison with the
analytical results shows that in this parameter range the BGK scattering
integral in the one-dimensional Boltzmann equation (\ref{Boltzmann equation})
adequately models the scattering in the 3D system.

The next step is to find the high-frequency conductivity in this nonequilibrium
state. We consider perturbations with frequency $\omega$ and wave-vector $k$ of
the form $E(x,t)=E_s+E_{\omega,k} e^{-i\omega t+ikx}$,
$f(x,\varphi,t)=f_s(\varphi)+f_{\omega,k}(\varphi) e^{-i\omega t+ikx}$ and
solve the linearized kinetic equation
\begin{equation}
  \frac{\partial f_{\omega,k}}{\partial \varphi}
  +i(\alpha+\kappa\sin\varphi)f_{\omega,k}= -\frac{E_{\omega,k}}{E_s}
  \frac{\partial f_s}{\partial \varphi}+\frac{\nu n_{\omega,k}}{\Omega n_0}f_0,
\end{equation}
with $$\alpha=-(\omega+i\nu)/\Omega,\ \ \ \
\kappa=\frac{\Delta d}{2\hbar\Omega}k.$$

For the differential conductivity we find (following the method developed in
\cite{Ignatov87})
\begin{equation}\label{sigma}
  \sigma(\omega,k)=\frac{j_{\omega,k}}{E_{\omega,k}}=
  \frac{2\pi\varepsilon_0e^2\Delta d}{\hbar\kappa}\sum_{m=-\infty}^\infty
  \frac{i^{m-1}mJ_m(\kappa)}{\omega+i\nu-m\Omega}
  \int_{-\pi}^\pi e^{-i\kappa\cos\varphi-im\varphi}\left[
  \frac{\partial f_s}{\partial\varphi}-i\frac{\nu\Omega\kappa}
  {\omega_p^2}f_0\right]d\varphi,
\end{equation}
where we introduced plasma frequency $\omega_p=\sqrt{\frac{e^2\Delta n_0d^2}
{\hbar^2\varepsilon'\varepsilon_0}}$, $\varepsilon'$ is the interlayer
dielectric constant, $\varepsilon_0$ is the vacuum dielectric constant. This is
the general expression for any nonequilibrium function $f_s(\varphi)$. With the
injection source (\ref{in-delta}) and $f_s(\varphi)$ given by (\ref{f_s}) we
obtain
$$\sigma(\omega,\kappa)=\sigma_E(\omega,\kappa)+\sigma'(\omega,\kappa),$$
\begin{equation}\label{IS sigma}
  \sigma_E(\omega,\kappa)=-\frac{\varepsilon'\varepsilon_0\nu\Omega}{I_0(\frac{\Delta}{2T})}
  \sum_{m,l=-\infty}^{\infty}\left(1-\frac{l\omega_p^2}{\kappa\Omega(\nu+il\Omega)}
  \right)\frac{mi^lI_l(\frac{\Delta}{2T})J_m(\kappa)J_{m-l}(\kappa)}
  {\omega+i\nu-m\Omega},
\end{equation}
\begin{equation}\label{hot sigma}
  \sigma'(\omega,\kappa)=\frac{\varepsilon'\varepsilon_0\Omega\eta\omega_p^2}{\kappa}
  \sum_{m,l=-\infty}^{\infty}\frac{lmi^lJ_m(\kappa)J_{m-l}(\kappa)}
  {(\nu+\eta\Omega+il\Omega)(\omega+i\nu-m\Omega)}
  \left[e^{-il\varphi'}-\frac{I_l\left(\frac{\Delta}{2T}\right)}{I_0\left(\frac{\Delta}{2T}\right)}
  \frac{\nu}{\nu+il\Omega}\right],
\end{equation}
$\sigma_E(\omega,\kappa)$ was obtained previously by Ignatov and Shashkin
\cite{Ignatov87}.

Now let us consider the conductivity in long-wavelength limit
$\sigma(\omega)=\sigma(\omega,\kappa\rightarrow 0)$.
Taking into account that $J_n(x)\rightarrow\frac{x^n}{2^nn!}$ for $x\rightarrow
0$ and  $n>0$, we obtain
\begin{equation}\label{sigma_E(omega)}
  \sigma_E(\omega)=
  \sigma_0\frac{I_1}{I_0}\frac{\nu^2(\Omega^2-\nu^2
  +i\nu\omega)}{(\nu^2+\Omega^2)[(\omega+i\nu)^2-\Omega^2]},
\end{equation}
which was first obtained by Ktitorov et al., and
\begin{equation}\label{sigma'(omega)}
\begin{array}{c} \displaystyle
  \sigma'(\omega)=\frac{\eta\sigma_0\nu\Omega}{(\tilde{\nu}^2+\Omega^2)[(\omega+i\nu)^2-\Omega^2]}
  \times \\
  \displaystyle
  \times\left\{(\Omega^2-\nu\tilde{\nu}+i\tilde{\nu}\omega)\cos\varphi'
  -\Omega(\omega+i\nu+i\tilde{\nu})\sin\varphi'
  -\frac{I_1}{I_0}
  \frac{\nu\Omega^2(\tilde{\nu}+\nu)+\nu(\tilde{\nu}\nu-\Omega^2)(i\omega-\nu)}{(\nu^2+\Omega^2)}
  \right\},
\end{array}
\end{equation}
where $\displaystyle\sigma_0=\frac{\varepsilon'\varepsilon_0\omega_p^2}{\nu}$,
$\tilde{\nu}=\nu+\eta\Omega$, the arguments $\Delta/2T$ of Bessel functions are
omitted.

The differential conductivity with hot electron injection is shown in
Fig.\,\ref{sigma-fig}(a) (solid lines) together with the conductivity without
injection (broken line). At strong enough injection (thick solid line) the hot
differential conductivity is significantly different from the cold one: now NDC
takes place above the Bloch frequency, and in the entire low-frequency part it
is positive. The results of the corresponding Monte-Carlo simulations at large
ac signal ($E_{\omega}=12$ kV/cm) are shown in Fig.\,\ref{sigma-fig}(b). They
confirm our main conclusion about the "inversion" of the differential
conductivity in a nonequilibrium state and demonstrate that at high fields the
effect can be even stronger.

Note, that in the regime with Bloch oscillations ($\Omega>\nu$) the important
parameter $\eta$ is determined by the competition between the pumping frequency
$\nu_t=Q/n_0$ and Bloch frequency $\Omega$. The desirable effect takes place
when $\eta$ is larger then some critical value $\eta_c$. Large $\eta$ can be
achieved at low temperatures with a low equilibrium density $n_0$. On the other
hand, at high (room) temperatures, when $T>\Delta$, $\eta_c$ becomes smaller,
because the equilibrium distribution function is flat in the miniband and
therefore the effect of pumping is more pronounced. That gives hope to observe
the effect at low as well as at room temperatures.

An important question is the choice of the injection momentum $p'$. Our
calculations show that the effect takes place if particles are injected at high
enough energies, $\epsilon_p>\Delta/2$. The injection with $p'$ and $-p'$ leads
qualitatively to the same result, but at positive $p'$ the effect is stronger
and exists in a larger frequency range. It can be easily seen from
Eq.\,(\ref{sigma'(omega)}) that $\Re\sigma'(\omega\rightarrow 0)\propto
-\cos\varphi'$ (at $\Omega^2>\nu(\nu+\eta\Omega)$) and
$\Re\sigma'(\omega\gg\Omega)\propto-\sin\varphi'$, such that the low-frequency
instability is suppressed equally by $p'$ and $-p'$ injection, but the
high-frequency instability with $\omega\gg\Omega$ exists only at $\varphi'>0$.

In conclusion, we have shown that hot particle injection into the upper part of
the miniband leads to a qualitative change of the frequency behaviour of the
differential conductivity. The region of NDC is shifted from low frequencies
(below the Bloch frequency) to higher frequencies. Consequently, the domain
instability is suppressed. To further support this result we considered the
instability of space-charge waves with finite wave vectors, the spectrum
$\omega(k)$ of which is given by the solution of the dispersion equation
\begin{equation}\label{eps}
  \varepsilon(\omega,k)=\varepsilon'+i\frac{\sigma(\omega,k)}{\varepsilon_0\omega}=0.
\end{equation}
In the high-field limit ($\Omega\gg\nu,\omega_p$) we found that a low-frequency
drift-diffusion mode with $\omega(k)\ll \Omega$, which is unstable without
injection, now becomes damped. This means, that the homogeneous state becomes
stable, and a superlattice in that state can be used as high-frequency
oscillator if we couple it to some external circuit (resonator or wave-guide).

The narrow hot particle source (\ref{in-delta}) may be realized by tunneling
from a metal or doped semiconductor \cite{Rauch97,Rauch98}, or by tunneling
from a narrow miniband superlattice. In the case of tunneling from the end of a
superlattice "from layer to layer" our approach is valid for an active region
with the length of the order of the Bloch length $l_B=\Delta/2eE_s\gg d$. The
consideration of an inhomogeneous distribution function in a real superlattice
with leads requires more sophisticated approaches \cite{Wacker02,Willenberg02},
which link Bloch oscillator and quantum cascade laser physics.

Qualitatively the same results are obtained for the optical excitation when
$S(p)\propto[\delta(p-p')+\delta(p+p')]$. The consideration of optical pumping
is the subject of a future publication. We think that the prospective way to
the practical realization of the considered effect is the combination of
injection (or optical excitation) of particles into the second miniband with
relaxation from the second miniband into the upper part of the first active
miniband.

We would like to thank V.Ya. Aleshkin,  A.A. Andronov, A.A. Ignatov, K.F. Renk,
and Yu.A. Romanov for discussions. This work is supported by Graduiertenkolleg
"Nichtlinearit\"{a}t und Nichtgleichgewicht in kondensierter Materie", Russian
Foundation for Basic Research grant 02-02-16385 and NATO Science program
"Science for Peace", project SfP-973799 (Semiconductors).


\begin{thebibliography}{16}
\expandafter\ifx\csname natexlab\endcsname\relax\def\natexlab#1{#1}\fi
\expandafter\ifx\csname bibnamefont\endcsname\relax
  \def\bibnamefont#1{#1}\fi
\expandafter\ifx\csname bibfnamefont\endcsname\relax
  \def\bibfnamefont#1{#1}\fi
\expandafter\ifx\csname citenamefont\endcsname\relax
  \def\citenamefont#1{#1}\fi
\expandafter\ifx\csname url\endcsname\relax
  \def\url#1{\texttt{#1}}\fi
\expandafter\ifx\csname urlprefix\endcsname\relax\def\urlprefix{URL }\fi
\providecommand{\bibinfo}[2]{#2} \providecommand{\eprint}[2][]{\url{#2}}

\bibitem[{\citenamefont{Esaki and Tsu}(1970)}]{Esaki70}
\bibinfo{author}{\bibfnamefont{L.}~\bibnamefont{Esaki}} \bibnamefont{and}
  \bibinfo{author}{\bibfnamefont{R.}~\bibnamefont{Tsu}}, \bibinfo{journal}{IBM
  J. Res. Dev.} \textbf{\bibinfo{volume}{14}}, \bibinfo{pages}{61}
  (\bibinfo{year}{1970}).

\bibitem[{\citenamefont{Ktitorov et~al.}(1972)\citenamefont{Ktitorov, Simin,
  and Sindalovskii}}]{Ktitorov72}
\bibinfo{author}{\bibfnamefont{S.}~\bibnamefont{Ktitorov}},
  \bibinfo{author}{\bibfnamefont{G.}~\bibnamefont{Simin}}, \bibnamefont{and}
  \bibinfo{author}{\bibfnamefont{V.}~\bibnamefont{Sindalovskii}},
  \bibinfo{journal}{Sov. Phys. -- Solid State}
  \textbf{\bibinfo{volume}{13}}(\bibinfo{number}{8}), \bibinfo{pages}{1872}
  (\bibinfo{year}{1972}), \bibinfo{note}{[Fiz. Tverd. Tel. {\bf 13}, 2230-2233
  (1971)]}.

\bibitem[{\citenamefont{Ignatov et~al.}(1993)\citenamefont{Ignatov, Renk, and
  Dodin}}]{Ignatov93}
\bibinfo{author}{\bibfnamefont{A.}~\bibnamefont{Ignatov}},
  \bibinfo{author}{\bibfnamefont{K.}~\bibnamefont{Renk}}, \bibnamefont{and}
  \bibinfo{author}{\bibfnamefont{E.}~\bibnamefont{Dodin}},
  \bibinfo{journal}{Phys. Rev. Lett.} \textbf{\bibinfo{volume}{70}},
  \bibinfo{pages}{1996} (\bibinfo{year}{1993}).

\bibitem[{\citenamefont{B\"{u}ttiker and Thomas}(1977)}]{Buttiker77-79}
\bibinfo{author}{\bibfnamefont{M.}~\bibnamefont{B\"{u}ttiker}}
  \bibnamefont{and} \bibinfo{author}{\bibfnamefont{H.}~\bibnamefont{Thomas}},
  \bibinfo{journal}{Phys. Rev. Lett.} \textbf{\bibinfo{volume}{38}},
  \bibinfo{pages}{78} (\bibinfo{year}{1977}), \bibinfo{note}{\uppercase{Z}.
  Phys. {\bf 33}, 275 (1979), Z. Phys. {\bf 34}, 301 (1979)}.

\bibitem[{\citenamefont{Schomburg et~al.}(1998)\citenamefont{Schomburg,
  Blomeier, Hofbeck, Grenzer, Brandl, Lingott, Ignatov, Renk, Pavel'ev,
  Koschurinov et~al.}}]{Schomburg98}
\bibinfo{author}{\bibfnamefont{E.}~\bibnamefont{Schomburg}},
  \bibinfo{author}{\bibfnamefont{T.}~\bibnamefont{Blomeier}},
  \bibinfo{author}{\bibfnamefont{K.}~\bibnamefont{Hofbeck}},
  \bibinfo{author}{\bibfnamefont{J.}~\bibnamefont{Grenzer}},
  \bibinfo{author}{\bibfnamefont{S.}~\bibnamefont{Brandl}},
  \bibinfo{author}{\bibfnamefont{I.}~\bibnamefont{Lingott}},
  \bibinfo{author}{\bibfnamefont{A.}~\bibnamefont{Ignatov}},
  \bibinfo{author}{\bibfnamefont{K.}~\bibnamefont{Renk}},
  \bibinfo{author}{\bibfnamefont{D.}~\bibnamefont{Pavel'ev}},
  \bibinfo{author}{\bibfnamefont{Y.}~\bibnamefont{Koschurinov}},
  \bibnamefont{et~al.}, \bibinfo{journal}{Phys. Rev. B}
  \textbf{\bibinfo{volume}{58}}(\bibinfo{number}{7}), \bibinfo{pages}{4035}
  (\bibinfo{year}{1998}).

\bibitem[{\citenamefont{Schomburg et~al.}(1999)\citenamefont{Schomburg, Henini,
  Chamberlain, Steenson, Brandl, Hofbeck, Renk, and Wegscheider}}]{Schomburg99}
\bibinfo{author}{\bibfnamefont{E.}~\bibnamefont{Schomburg}},
  \bibinfo{author}{\bibfnamefont{M.}~\bibnamefont{Henini}},
  \bibinfo{author}{\bibfnamefont{J.}~\bibnamefont{Chamberlain}},
  \bibinfo{author}{\bibfnamefont{D.}~\bibnamefont{Steenson}},
  \bibinfo{author}{\bibfnamefont{S.}~\bibnamefont{Brandl}},
  \bibinfo{author}{\bibfnamefont{K.}~\bibnamefont{Hofbeck}},
  \bibinfo{author}{\bibfnamefont{K.~F.} \bibnamefont{Renk}}, \bibnamefont{and}
  \bibinfo{author}{\bibfnamefont{W.}~\bibnamefont{Wegscheider}},
  \bibinfo{journal}{Appl. Phys. Lett.}
  \textbf{\bibinfo{volume}{74}}(\bibinfo{number}{15}), \bibinfo{pages}{2179}
  (\bibinfo{year}{1999}).

\bibitem[{\citenamefont{Romanov et~al.}(1978)\citenamefont{Romanov, Bovin, and
  Orlov}}]{Romanov78}
\bibinfo{author}{\bibfnamefont{Y.}~\bibnamefont{Romanov}},
  \bibinfo{author}{\bibfnamefont{V.}~\bibnamefont{Bovin}}, \bibnamefont{and}
  \bibinfo{author}{\bibfnamefont{L.}~\bibnamefont{Orlov}},
  \bibinfo{journal}{Sov. Phys. Semicond.}
  \textbf{\bibinfo{volume}{12}}(\bibinfo{number}{9}), \bibinfo{pages}{987}
  (\bibinfo{year}{1978}), \bibinfo{note}{[Fiz. Tekh. Poluprovodn. {\bf 12},
  1665 (1978)]}.

\bibitem[{\citenamefont{Kroemer}(2000)}]{Kroemer00}
\bibinfo{author}{\bibfnamefont{H.}~\bibnamefont{Kroemer}},
  \bibinfo{journal}{cond-mat/0009311}  (\bibinfo{year}{2000}).

\bibitem[{\citenamefont{Romanov and Romanova}(2000)}]{Romanov00}
\bibinfo{author}{\bibfnamefont{Y.}~\bibnamefont{Romanov}} \bibnamefont{and}
  \bibinfo{author}{\bibfnamefont{Y.}~\bibnamefont{Romanova}},
  \bibinfo{journal}{J. Exp. Theor. Phys. (Russia)}
  \textbf{\bibinfo{volume}{91}}(\bibinfo{number}{5}), \bibinfo{pages}{1033}
  (\bibinfo{year}{2000}), \bibinfo{note}{[Zh. Eksp. Teor. Fiz. {\bf 118}, 1193
  (2000)]}.

\bibitem[{\citenamefont{Cannon et~al.}(2000)\citenamefont{Cannon, Kusmartsev,
  Alekseev, and Campbell}}]{Cannon00}
\bibinfo{author}{\bibfnamefont{E.~H.} \bibnamefont{Cannon}},
  \bibinfo{author}{\bibfnamefont{F.~V.} \bibnamefont{Kusmartsev}},
  \bibinfo{author}{\bibfnamefont{K.~N.} \bibnamefont{Alekseev}},
  \bibnamefont{and} \bibinfo{author}{\bibfnamefont{D.~K.}
  \bibnamefont{Campbell}}, \bibinfo{journal}{Phys. Rev. Lett.}
  \textbf{\bibinfo{volume}{85}}(\bibinfo{number}{6}), \bibinfo{pages}{1302}
  (\bibinfo{year}{2000}).

\bibitem[{\citenamefont{Ignatov and Shashkin}(1987)}]{Ignatov87}
\bibinfo{author}{\bibfnamefont{A.}~\bibnamefont{Ignatov}} \bibnamefont{and}
  \bibinfo{author}{\bibfnamefont{V.}~\bibnamefont{Shashkin}},
  \bibinfo{journal}{Sov. Phys. JETP}
  \textbf{\bibinfo{volume}{66}}(\bibinfo{number}{3}), \bibinfo{pages}{526}
  (\bibinfo{year}{1987}), \bibinfo{note}{[Zh. Eksp. Teor. Fiz. {\bf 93}, 935
  (1987)]}.

\bibitem[{\citenamefont{Jacoboni and Reggiani}(1983)}]{Jacoboni83}
\bibinfo{author}{\bibfnamefont{C.}~\bibnamefont{Jacoboni}} \bibnamefont{and}
  \bibinfo{author}{\bibfnamefont{L.}~\bibnamefont{Reggiani}},
  \bibinfo{journal}{Rev. of Mod. Phys.}
  \textbf{\bibinfo{volume}{55}}(\bibinfo{number}{3}), \bibinfo{pages}{645}
  (\bibinfo{year}{1983}).

\bibitem[{\citenamefont{Rauch et~al.}(1997)\citenamefont{Rauch, Strasser,
  Unterrainer, Gornik, and Brill}}]{Rauch97}
\bibinfo{author}{\bibfnamefont{C.}~\bibnamefont{Rauch}},
  \bibinfo{author}{\bibfnamefont{G.}~\bibnamefont{Strasser}},
  \bibinfo{author}{\bibfnamefont{K.}~\bibnamefont{Unterrainer}},
  \bibinfo{author}{\bibfnamefont{E.}~\bibnamefont{Gornik}}, \bibnamefont{and}
  \bibinfo{author}{\bibfnamefont{B.}~\bibnamefont{Brill}},
  \bibinfo{journal}{Appl. Phys. Lett.}
  \textbf{\bibinfo{volume}{70}}(\bibinfo{number}{5}), \bibinfo{pages}{649}
  (\bibinfo{year}{1997}).

\bibitem[{\citenamefont{Rauch et~al.}(1998)\citenamefont{Rauch, Strasser,
  Unterrainer, Boxleitner, Gornik, and Wacker}}]{Rauch98}
\bibinfo{author}{\bibfnamefont{C.}~\bibnamefont{Rauch}},
  \bibinfo{author}{\bibfnamefont{G.}~\bibnamefont{Strasser}},
  \bibinfo{author}{\bibfnamefont{K.}~\bibnamefont{Unterrainer}},
  \bibinfo{author}{\bibfnamefont{W.}~\bibnamefont{Boxleitner}},
  \bibinfo{author}{\bibfnamefont{E.}~\bibnamefont{Gornik}}, \bibnamefont{and}
  \bibinfo{author}{\bibfnamefont{A.}~\bibnamefont{Wacker}},
  \bibinfo{journal}{Phys. Rev. Lett.}
  \textbf{\bibinfo{volume}{81}}(\bibinfo{number}{16}), \bibinfo{pages}{3495}
  (\bibinfo{year}{1998}).

\bibitem[{\citenamefont{Wacker}(2002)}]{Wacker02}
\bibinfo{author}{\bibfnamefont{A.}~\bibnamefont{Wacker}},
  \bibinfo{journal}{Phys. Rev. B} \textbf{\bibinfo{volume}{66}},
  \bibinfo{pages}{085326} (\bibinfo{year}{2002}).

\bibitem[{\citenamefont{Willenberg et~al.}(2002)\citenamefont{Willenberg,
  D\"{o}hler, and Faist}}]{Willenberg02}
\bibinfo{author}{\bibfnamefont{H.}~\bibnamefont{Willenberg}},
  \bibinfo{author}{\bibfnamefont{G.}~\bibnamefont{D\"{o}hler}},
  \bibnamefont{and} \bibinfo{author}{\bibfnamefont{J.}~\bibnamefont{Faist}},
  \bibinfo{journal}{cond-mat/0205359}  (\bibinfo{year}{2002}).

\end{thebibliography}
\end{document}